\pdfoutput=1
\documentclass[aps,prl,twocolumn,superscriptaddress,showpacs,amssymb,reprint,amsmath,preprintnumbers]{revtex4-1}

\usepackage{graphicx}
\usepackage{dcolumn}
\usepackage{bm}

\usepackage{afterpage}
\usepackage{natbib}
\begin{document}

\title{Spin density wave instability in a ferromagnet}


\author{Yan Wu}
\affiliation{Department of Physics and
Astronomy, Louisiana State University, Baton Rouge, LA 70803}

\author{Zhenhua Ning} 
\affiliation{Department of Physics and Astronomy,
Louisiana State University, Baton Rouge, LA 70803}

\author{Huibo Cao}
\affiliation{Quantum Condensed Matter Division, Oak Ridge National Laboratory,
Oak Ridge, TN 37831}

\author{Guixin Cao}
\affiliation{Department of Physics and Astronomy, Louisiana State
University, Baton Rouge, LA 70803}

\author{K. A. Benavides}
\affiliation{Department of Chemistry and Biochemistry,
The University of Texas at Dallas, Richardson, TX 75080}

\author{Gregory T.  McCandless}
\affiliation{Department of Chemistry and Biochemistry,
The University of Texas at Dallas, Richardson, TX 75080}

\author{R. Jin}
\affiliation{Department of Physics and Astronomy, Louisiana State
University, Baton Rouge, LA 70803}

\author{Julia Y. Chan}
\affiliation{Department of Chemistry and Biochemistry,
The University of Texas at Dallas, Richardson, TX 75080}

\author{W. A. Shelton}
\affiliation{Department of Chemical Engineering, Louisiana
State University, Baton Rouge, LA 70803}

\author{J. F. DiTusa}
\email[]{ditusa@phys.lsu.edu}
\affiliation{Department of Physics and Astronomy, Louisiana State
University, Baton Rouge, LA 70803}

\date{\today}
\begin{abstract}
Ferromagnetic (FM) and incommensurate spin-density wave (ISDW) states
are an unusual set of competing magnetic orders that are seldom
observed in the same material without application of a polarizing
magnetic field. We report, for the first time, the discovery of an
ISDW state that is derived from a FM ground state through a Fermi
surface (FS) instability in Fe$_3$Ga$_4$. This was achieved by
combining neutron scattering experiments with first principles
simulations. Neutron diffraction demonstrates that Fe$_3$Ga$_4$ is in
an ISDW state at intermediate temperatures and that there is a
conspicuous re-emergence of ferromagnetism above 360 K. First
principles calculations show that the ISDW ordering wavevector is in
excellent agreement with a prominent nesting condition in the
spin-majority FS demonstrating the discovery of a novel instability
for FM metals; ISDW formation due to Fermi surface nesting in a
spin-polarized Fermi surface.
\end{abstract}


\maketitle
Incommensurate spin density wave (ISDW) phases have long been known to
be generated from paramagnetic (PM) metals either through a Fermi
surface (FS) instability known as nesting or through the coupling of
local magnetic moments via the Ruderman-Kittel-Kasuya-Yosida (RKKY)
interaction\cite{overhauser,overhauser1}. Interestingly, no ISDW phase
stemming from an instability in the spin polarized FS of a
ferromagnetic (FM) metal has been previously identified. We have discovered
such a material in the transition metal binary compound Fe$_3$Ga$_4$
that hosts both FM and ISDW phases at zero applied field. Here, both
of these phases are stable over wide temperature ($T$) ranges with the FM
phase evident at temperatures above and below the $T$ range of
stability of the ISDW phase. Furthermore, we have discovered the
mechanism for the formation of the ISDW state in Fe$_3$Ga$_4$, which
is distinct from any previously identified and that makes clear the
connection between the FM and ISDW states. This mechanism, indicated
by our data and simulations, is likely not limited to Fe$_3$Ga$_4$,
suggesting that ISDW order or spin fluctuations stemming from
spin-polarized FS instabilities should be investigated in a wide range
of complex magnetic materials.

ISDW phases caused by FS instabilities are common in {\it d}-electron
systems such as the prototypical ISDW system
Cr\cite{fawcett,fawcett1}, while the local moment picture is usually
associated with magnetism in rare earth materials where the magnetic
moments derive from well localized {\it
f}-electrons\cite{barandiaran,islam,budko,good,lynn}. ISDW phases are
commonly associated with antiferromagnetic (AFM) interactions, as
opposed to incommensurate helical states which are more aptly
described as being derived from FM states. In fact, transitions
between ISDW and FM states in zero applied field are extraordinarily
rare in contrast to helical magnets where transitions to FM ground
states are common\cite{moral1975,mendive}. There are only a handful of
FM materials that display ISDW phases over narrow $T$ ranges with
distinct, fairly restrictive mechanisms. These include: ferromagnets
which seemingly avoid quantum criticality via a transition to a small
ordered moment ISDW, Nb$_{1-y}$Fe$_{2+y}$, YbRh$_2$Si$_2$, and
PrPtAl\cite{rauch,niklowitz,lausberg,abdul}; a frustrated magnetic
insulator described as a Kagom\'{e} staircase system that undergoes
transitions between several magnetic states,
Co$_3$V$_2$O$_8$\cite{ychen}; and a very small number of Ce and U
compounds which display ISDW states over a very narrow $T$ range just
above their FM transitions and where the mechanism is unclear,
CeRu$_2$Al$_2$B and UCu$_2$Si$_2$\cite{bhattacharyya,honda}. The
paucity of examples that we are able to find in the literature
stemming from decades of intense investigation of magnetic materials
is surprising.

In this letter, we focus on the investigation of the magnetic and
electronic structure of Fe$_3$Ga$_4$ because previous magnetization,
$M$, measurements indicated that there may be an interesting and
unusual re-entrant FM phase surrounding an AFM phase at both low and
high $T$\cite{mendez2015}. Neither the magnetic structure, nor the
character of the magnetic phases had been identified. The transitions
between FM and AFM phases in Fe$_3$Ga$_4$ can be tuned via $T$,
magnetic field, and
pressure\cite{mendez2015,wagini1966,duijn1999,duijnthesis}. We
discover through neutron scattering measurements that Fe$_3$Ga$_4$
adopts an ISDW ordering between 60 and 360 K. This magnetic structure
is nearly FM with magnetic moments being locally aligned but having a
sinusoidal amplitude modulation occurring over several crystalline
unit cells. In addition, we confirm the re-emergence of the FM phase
above 360 K so that this material appears to display a
thermodynamically odd re-entrant ferromagnetism\cite{mendez2015}.  A
clue as to the cause for this unusual behavior comes from simulations
of the electronic structure. In the non-magnetic case, our
calculations do not indicate regions of FS nesting. However, we
discovered significant nesting in the FM majority-spin FS with a
nesting vector that is in excellent agreement with the measured ISDW
wavevector, $q=$ (0 0 0.29).  The FS instability along with anomalies
in the charge transport support an itinerant mechanism for the
magnetism in Fe$_3$Ga$_4$. This indicates that the ISDW emerges from
an unstable FS in the {\it FM phase} so that a return to
ferromagnetism is required prior to reaching the high $T$ PM
phase. This mechanism, a FS instability in the majority band of a
ferromagnet that drives a transition to an ISDW, has, to our
knowledge, not been previously considered.

\begin{figure}[htb]
\includegraphics[width=1.0\linewidth,bb=195 85 475
 530,clip]{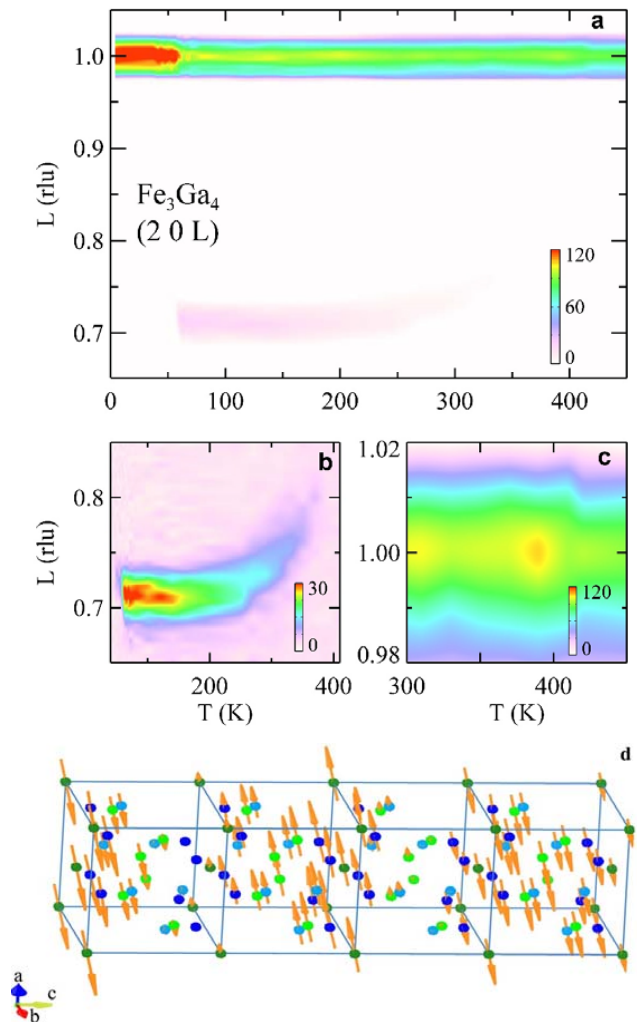}
\caption{\label{fig:contourplot} Temperature, $T$, dependence of the
neutron scattering intensity along the (2 0 $L$) direction. Scans were
performed in increments in $L$ of 0.005 reciprocal lattice units (rlu)
for $5 \le T \le 450$ K in increments of 5 K. Color-bars indicate the
scattering intensity in counts/s. Intensity plots displaying
scattering a) over the full $q$ and $T$ range of the data, b) near the
incommensurate wave-vector (2 0 1-$\delta$) for $5 \leq$$T$$\leq 450$
K, and c) in proximity to (2 0 1) for $300\leq T \leq 450$ K. d)
Magnetic structure at 100 K over a four unit cell length along the
$c$-axis depicting the ISDW state. Solid circles represent the 4
different Fe sites within the Fe$_3$Ga$_4$ unit
cell{\protect{\cite{suppmat}}}. }
\end{figure}

\begin{figure}[ht]
\includegraphics[width=1.0\linewidth,bb=60 365 575
 760,clip]{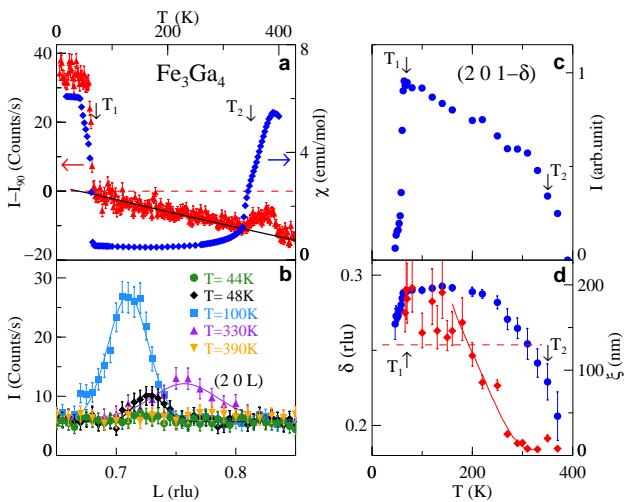}
\caption{\label{fig:FMorderingwithSus} Temperature dependent magnetic
scattering and magnetic susceptibility. (a) Neutron scattering
intensity at the (2 0 1) Bragg peak position, $I-I_{90}$, (red triangles)
and magnetic susceptibility, $\chi$, at 100 Oe (blue diamonds) vs.\
$T$. $I-I_{90}$ is the intensity, $I$, measured after subtraction of
the intensity at $90$ K, $I_{90}$. Solid line is a fit of the Debye
model\cite{DWmodel} to the data between 90 and 300 K with $\Theta_D$=
125 K\cite{mendez2015}. (b) Scattering along the (2 0 $L$) direction
at several representative temperatures, $T$'s, demonstrating the
incommensurate scattering. (c) Integrated intensity of the scattering
at (2 0 $1-\delta$) vs.\ $T$. (d) $T$ dependence of $\delta$ (blue
bullets) and the correlation length of the incommensurate scattering,
$\xi$, (red diamonds) vs. $T$. Solid line is a guide to the
eye. Dashed line is the instrumental resolution. Transition
temperatures $T_1$ and $T_2$ are indicated in frames a, c, and d. }
\end{figure}

Fe$_3$Ga$_4$ forms in a monoclinic ($C2/m$) crystal structure with no
evidence of structural change associated with the magnetic phase
transitions\cite{mendez2015,duijn1999,duijnthesis,philippe1,philippe2}.
The description of the crystal structure and an outline of our
experimental methods can be found in \cite{suppmat}. A summary of our
neutron scattering data is presented in Fig.~\ref{fig:contourplot}a,
which displays the intensity map for the scattering in the (2 0 $L$)
reciprocal lattice direction.  Three features of this plot represent
our main experimental findings. First, the large increase in the
scattering cross section at the (201) Bragg peak position below $\sim
60$ K demonstrates FM ordering consistent with the magnetic
susceptibility, $\chi$ (Fig.~\ref{fig:FMorderingwithSus}a), previous
measurements of $M$, and the M$\ddot{o}$ssbauer spectrum of
Fe$_3$Ga$_4$\cite{mendez2015,duijnthesis,duijn1999,Mossbauer1986,kawamiya2}.
Second, for $T$ just above 60 K, the reduced scattering at the Bragg
peak position is accompanied by an increase in scattering centered at
(2 0 0.71) indicating the existence of an incommensurate magnetic
phase.  The $T$ and $q$ dependence of this scattering contribution can
be better viewed in Fig.~\ref{fig:contourplot}b and in
Fig.~\ref{fig:FMorderingwithSus}b and c where an abrupt loss of
scattering is evident below 60 K while a more continuous change is
seen above 300 K. At these higher temperatures, the scattering moves
to somewhat larger $q$ before this signal is reduced below the
background at $T>360$ K. Third, in this same $T$ range, the scattering
at the (2 0 1) Bragg peak position increases
(Fig.~\ref{fig:contourplot}a, c and Fig. 2a) before decreasing at
$T>420$ K. This is consistent with $\chi$ which indicates a PM state
above 420 K\cite{mendez2015}. Thus, our data indicate a FM ordered
state below $T_1=60$ K transitioning to an incommensurate magnetic
state for $T_1 < T < T_2 = 360$ K along with a re-emergence of
ferromagnetism between $T_2$ and $T_3=420$ K.

Surveys of a large number of reciprocal lattice positions have allowed
a full refinement of the magnetic structure, the results of which are
presented in Table S2 for the low $T$ FM state and Table S3 for the
AFM order at 100 K in \cite{suppmat}. In Table S2\cite{suppmat}, we
report the size of the magnetic moment on each of the Fe sites refined
with the magnetic moment constrained to lie along the crystallographic
c-axis. These are comparable to the results of previously published
M$\ddot{o}$ssbauer experiments\cite{Mossbauer1986}, which are included
in the table, and are in good agreement with the average magnetic
moment determined from
$M$\cite{mendez2015,duijnthesis,Mossbauer1986}. When the constraint is
removed a small contribution along the $a$ and $b$ directions is found
that is smaller than the error in the refinement ($\sim0.3$
$\mu_B$). In Fig.~\ref{fig:FMorderingwithSus}a we present a comparison
of the scattering intensity, after subtraction of the Bragg scattering
at 90 K, to $\chi$, measured on the same crystal. The two regions in
$T$ where $\chi$ is large correspond well with the increased
scattering at (2 0 1).

The result of the refinement of the magnetic scattering at 100 K (AFM
phase) is significantly more complex with the best fit to our data
consisting of a structure that closely resembles that of an ISDW
having a propagation vector of $q=$ (0 0 0.29). After establishing the
ISDW character of the magnetic state, we have further constrained the
model such that crystallographically equivalent sites were required to
have the same magnetic moment amplitude. The results of this
refinement are displayed in Fig.~\ref{fig:contourplot}d. In Table
S3\cite{suppmat} the magnitude of the magnetic moments in one
particular unit cell is presented, but we note that the incommensurate
nature of the ordering results in moments that vary from cell-to-cell
accordingly. The refinement places the magnetic moments mostly along
the a-axis with a maximum magnitude of 2.31(6) $\mu_B$. However, there
is also a considerable contribution along the b- and c-axis of nearly
equal magnitude ($m_{max}$=0.58(6) $\mu_B$). Thus, there is a
significant non-collinear and non-coplanar magnetic moment indicated
by our data. Previously, we had speculated about such a non-coplanar
magnetic moment based upon a large topological Hall effect in the
range of temperatures and fields where the AFM is
stable\cite{mendez2015} and the data presented here confirm this
interpretation. 

The magnetic state represented in Fig.~\ref{fig:contourplot}d retains
a FM orientation of neighboring magnetic moments with an amplitude
that is modulated by the wave vector $\delta=0.29$ in reciprocal
lattice units. This is distinct from the simpler ISDW materials such
as Cr\cite{fawcett} where neighboring sites are AFM aligned. Thus, the
ISDW state appears to be a long wavelength modulation of a FM state.
In fact, the average size of the magnetic moments in the ISDW state
are within error of those in the FM state indicating a conservation of
total moment magnitude at $T_1$. However, there is a significant
change in the direction of the magnetic moments in the FM and AFM
structures. The complexity of the ISDW phase and the significant
differences in the two competing magnetic phases highlight the
question of the mechanism driving the ISDW ordering and its
instability at low $T$.

The evolution of $\delta$ and the correlation length for the magnetic
scattering, $\xi$, with $T$ can lend insight into the character of the
magnetic state and the competing interactions which are of clear
importance in Fe$_3$Ga$_4$\cite{suppmat}. In
Figs.~\ref{fig:contourplot}b, \ref{fig:FMorderingwithSus}b and
\ref{fig:FMorderingwithSus}d the $T$ evolution of $\delta$ is
displayed. This $T$ dependence is usually associated with the
competition between the electronic degrees of freedom responsible for
the incommensurate nature of the wavevector and the lattice degrees of
freedom which prefer a commensurate density wave\cite{feng2015}. Here,
we observe a decrease in $\delta$ as the FM phases are approached
either by warming or cooling with a large decrease in $\delta$
apparent above 250 K. Preceding this decrease in $\delta$ is a
precipitous reduction in $\xi$ as determined from the widths of the (2
0 1-$\delta$) scattering peak for $T> 200$ K
(Fig.~\ref{fig:FMorderingwithSus}a and d). Above 300 K $\xi$ has
decreased such that it is equivalent to several wavelengths of the
ISDW phase ($2\pi/\delta$). Thus, there is an extended $T$ range where
only short range ordering exists.

\begin{figure}[ht]
\includegraphics[angle=0,width=0.8\linewidth,bb=70 355 555
 690,clip]{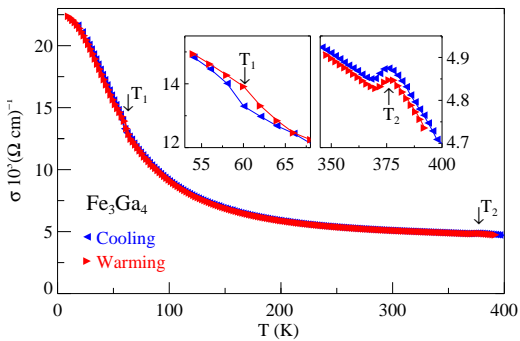}
\caption{\label{fig:conductivity} Temperature, $T$, dependence of the
electrical conductivity conductivity, $\sigma$, measured with current
along the c-axis. Insets: changes to $\sigma$ at the magnetic phase
transitions.  Transition temperatures $T_1$ and $T_2$ are indicated.
}
\end{figure}

Independent of the mechanism responsible, most ISDW phases are
accompanied by a discontinuous decrease in conductivity, $\sigma$, and
an increase in the Hall constant as the result of a partial FS
gapping. This is a consequence of either FS nesting\cite{overhauser1}
or simply the additional periodicity associated with the ISDW phase.
Fe$_3$Ga$_4$ does not disappoint in this regard as we identify both a
decreased $\sigma(T)$ at $T_1$ and a somewhat smaller increase at
$T_2$ (Fig.~\ref{fig:conductivity}) indicating a change to the
FS\cite{duijnthesis,mendez2015}. Furthermore, the ordinary Hall
coefficient is observed to undergo large changes at both $T_1$ and
$T_2$\cite{mendez2015}.

\begin{figure}[ht]
\includegraphics[angle=0,width=0.7\linewidth,bb=60 55 480
 490,clip]{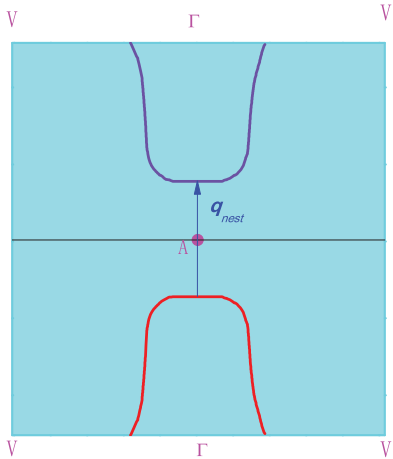}
\caption{\label{fig:estruct}Calculated Fermi surface of Fe$_3$Ga$_4$
on a (010) plane through $\Gamma$ with the proposed nesting
illustrated. Here $\Gamma$=(0,0,0), $A$=(0,0,1/2) and
$V$=(1/2,0,0). Red and violet are sheets of Fermi surface belonging
to adjacent Brillouin zones.}
\end{figure}

To establish the itinerant nature of the magnetism, the electronic
structure of Fe$_3$Ga$_4$ was calculated in the nonmagnetic and spin
polarized phases employing the full potential linearized augmented
plane wave method\cite{FLAPW}. The resulting Fermi Surfaces were
examined for possible nesting along the $c$-axis. Although we find no
such nesting condition in the non-magnetic FS, the majority
spin-band FS in the FM state contains a FS sheet with a substantial
region that is flat and perpendicular to the $c^{*}$ direction (see
Fig.~\ref{fig:estruct}). Here, the nesting across the Brillouin zone
boundary is demonstrated indicating the likely formation of density
wave ordering. The nesting wavevector shown in this figure, $q_{nest}=
$(0 0 0.276), is in very good agreement with the experimentally
observed ISDW providing compelling evidence that the FM state is
unstable to the formation of a density wave.

We have carried out an extensive neutron diffraction investigation of
Fe$_3$Ga$_4$ solving, for the first time, the magnetic structure of
the AFM state finding an ISDW phase. The ground state is confirmed as
a robust FM state and the re-emergence of this ferromagnetism is also
confirmed for $360 \le T \le 420$ K. Although the magnetic
contribution to the neutron scattering signal is small and data too
sparse above 360 K to make a convincing comparison between the low and
high $T$ FM states, our $\chi(T)$ and $M(H,T)$ data display identical
anisotropy as well as similar moment magnitudes and field dependencies
indicating that these FM states are likely
identical\cite{mendez2015}. Furthermore, electronic structure
calculations reveal a FS sheet in the majority band of the FM state
that is unstable towards density wave formation with a nesting vector
that matches the neutron data both in direction and
magnitude. Interestingly, the ISDW state likely results from a FS
instability in only one of the spin-polarized bands of Fe$_3$Ga$_4$, a
mechanism that, thus far, has not been considered or identified in
other materials. The details of how the transition between itinerant
FM and ISDW states takes place are not yet clear since a residual
magnetic moment may be expected for a simple FS reconstruction in
conflict with our $M(H)$\cite{mendez2015} and neutron diffraction
data.

These conclusions place Fe$_3$Ga$_4$ as a unique compound among the
large number of FM materials reported in the literature such that
identifying materials that are more than tangentially related is
difficult. Extending the conversation to examples of FM materials with
consequential FS nesting in their PM states broadens the number of
comparisons slightly as this is only somewhat rare. Most of these are
associated with competing orders and none, that we are aware of,
report ISDW states resulting from nesting. Competing orders occur, for
example, in SmNiC$_2$ which has a charge density wave (CDW) phase that
exists only above its Curie $T$\cite{Lei}. In addition, nesting of the
PM FS is common in the heavy rare earth elemental metals which tend to
have FM ground states\cite{moral1975,mendive}. This nesting leads to
transitions from PM into helimagnetic, rather than ISDW, states. More
interesting is the idea that triplet superconductivity in FM UGe$_2$
is related to coupled CDW-ISDW fluctuations that emerge with pressure
within the FM phase\cite{watanabe}.

We are left with two important unanswered questions about the
mechanism we discovered for the transitions between FM and ISDW states
in Fe$_3$Ga$_4$. The first is the character of the magnetism and its
relationship to the conducting charge carriers. Somewhat in contrast
to an itinerant magnetism in Fe$_3$Ga$_4$ are the large magnetic
moments evident in $M$, M$\ddot{o}$ssbauer spectra, simulation, and
the neutron scattering cross-sections suggesting magnetic moments may
be more localized to the Fe sites. This would indicate an important
coupling mechanism between the charge carriers and the more localized
electrons responsible for the magnetic moments.  A similar cooperative
mechanism has been proposed for GdSi where both RKKY and FS nesting
are thought to play a role in creating an ISDW state\cite{feng2013a}
and for the Fe-based superconducting families where this coupling is
responsible for the various AFM
structures\cite{hosono,paglione,stewart}. The second related question
is the cause of the FM ground state. We speculate that the
re-emergence of FM ground state at low $T$ may be a result of a loss
of RKKY coupling with the establishment of partial energy gaps in the
FS of the ISDW state, as has been suggested for the helimagnetic rare
earth elemental metals such as Dy and
Tb\cite{moral1975,mcewen1978,jensen1991,mendive}. In these materials
an abrupt transition from FM-to-helimagnetic ordering results from
superzone gaps in the FS of the incommensurate phase along with
considerations of spin-orbit coupling\cite{moral1975}. As such, the
incommensurate phase appears to be self limiting as the ordering
creates a partial energy gap that removes carriers responsible for the
RKKY coupling.  Additionally, there is a complex and interesting
variation of the magnetic wavevector and metamagnetic fields that is
in many ways similar to what we observe in Fe$_3$Ga$_4$. Despite these
remaining issues, our data and computational results indicate that
Fe$_3$Ga$_4$ is the first material that has been discovered to evolve
from a FM FS to an ISDW, a discovery that suggests that there is
likely a rich and mostly unrecognized competition between magnetic
states caused by FS instabilities of spin polarized bands.

\begin{acknowledgments}
This work was supported by the U.S. Department of Energy under EPSCoR
Grant No. DE-SC0012432 with additional support from the Louisiana
Board of Regents. The X-ray diffraction experiments were supported by
the National Science Foundation under grant number DMR-1360863 (JYC).
\end{acknowledgments}
\nocite{*}
\bibliography{Fe3Ga4JFD_65}

\end{document}


\preprint{APS/123-QED} 
\title{Supplementary Materials for Spin density wave instability in a ferromagnet}


\author{Yan Wu}
\affiliation{Department of Physics and
Astronomy, Louisiana State University, Baton Rouge, LA 70803}

\author{Zhenhua Ning} 
\affiliation{Department of Physics and Astronomy,
Louisiana State University, Baton Rouge, LA 70803}

\author{Huibo Cao}
\affiliation{Quantum Condensed Matter Division, Oak Ridge National Laboratory,
Oak Ridge, TN 37831}

\author{Guixin Cao}
\affiliation{Department of Physics and Astronomy, Louisiana State
University, Baton Rouge, LA 70803}

\author{K. A. Benavides}
\affiliation{Department of Chemistry and Biochemistry,
The University of Texas at Dallas, Richardson, TX 75080}

\author{Gregory T.  McCandless}
\affiliation{Department of Chemistry and Biochemistry,
The University of Texas at Dallas, Richardson, TX 75080}

\author{R. Jin}
\affiliation{Department of Physics and Astronomy, Louisiana State
University, Baton Rouge, LA 70803}

\author{Julia Y. Chan}
\affiliation{Department of Chemistry and Biochemistry,
The University of Texas at Dallas, Richardson, TX 75080}

\author{W. A. Shelton}
\affiliation{Department of Chemical Engineering, Louisiana
State University, Baton Rouge, LA 70803}

\author{J. F. DiTusa}
\email[]{ditusa@phys.lsu.edu}
\affiliation{Department of Physics and Astronomy, Louisiana State
University, Baton Rouge, LA 70803}

\date{\today}
\begin{abstract}
\end{abstract}
\maketitle Here we present a summary of our experimental methods
including detailed information about the crystal and magnetic
structure determined from the X-ray and neutron diffraction
experiments.

Fe$_3$Ga$_4$ forms in a monoclinic ($C2/m$) crystal structure with no
evidence of structural change associated with the magnetic phase
transitions\cite{mendez2015,duijn1999,duijnthesis,philippe1,philippe2}.
This structure is somewhat complex having 4 crystallographically
unique Fe sites and 4 unique Ga sites within the unit cell (Fig.\ 1d
and table S1.)

Single crystals of Fe$_3$Ga$_4$ with masses up to 100 mg were grown
from high purity Fe powder and gallium pieces that were melted in an
RF furnace prior to employing floating zone techniques.  The resulting
crystals were sealed in quartz tubes under vacuum and annealed at 550
$^{\circ}$C for 48 hours.

The structure of the crystals was determined by single crystal X-ray
diffraction using a Bruker D8 Quest Kappa single-crystal
diffractometer (Mo $K\alpha$ $I\mu S$ microfocus source,
$\lambda=0.71073$ \AA) operating at 50 kV and 1 mA, a HELIOS optics
monochromator, and a CMOS detector. These measurements confirmed a
base-centered monoclinic structure having the space group
$C2/m$\cite{philippe1,philippe2} for $300 \le T \le 380$ K.  Single
crystal X-ray diffraction data were collected at 300, 328, 350, and
380 K. Crystals were cut to an appropriate size and mounted onto a
glass fiber using epoxy. The Bruker program SADABS was used to correct
the collected data for absorption\cite{krause}. The intrinsic phasing
method in SHELXT\cite{sheldrick} was used to obtain starting
crystallographic models and SHELXL2014\cite{sheldrick2} was used to
refine atomic sites. The crystallographic parameters and atomic
positions for Fe$_3$Ga$_4$ at 300 K are provided in Table S1 to serve
as a representative of all the data collected.

\begin{table*}[hbt]\centering
\begin{tabular}{|c|c|c|c|c|c|}
\hline \multicolumn{1}{|c|}{\bf Label} &\multicolumn{1}{c|}{\textbf {Site Symm.}} &
\multicolumn{1}{c|}{\textbf {x}} & \multicolumn{1}{c|}
{\textbf {y}} & \multicolumn{1}{c|}
{\textbf {z}} & \multicolumn{1}{c|}{\textbf {$U_{eq}($\AA$^2)^a$}}  \\ %
\hline \centering
Fe1&2a&0&0&0&0.00314(18)\\ 
Fe2&4i&0.49454(7)&0&0.31074(9)&0.00373(14)\\ 
Fe3&4i&0.22477(7)&0&0.63799(9)&0.00320(14)\\ 
Fe4&8j&0.13570(5)&0.20036(6)&-0.15173(6)&0.00356(12)\\ 
Ga1&4i&0.26982(6)&0&0.08459(7)&0.00667(13)\\ 
Ga2&4i&0.04142(6)&0&0.34795(7)&0.00608(13)\\ 
Ga3&8j&0.60842(4)&0.20288(5)&0.15120(5)&0.00586(11)\\ 
Ga4&8j&0.35199(4)&0.18656(5)&0.44954(5)&0.00703(11)\\
\hline
\hline
\end{tabular}
\caption{\label{table:xraydata} Atomic Positions of Fe$_3$Ga$_4$ at
300 K determined from single crystal X-ray diffraction: Space Group
$C2/m$, $a$ = 10.0949(16) \AA, b = 7.6603(16) \AA, c = 7.8703(14) \AA,
$\beta$ = 106.286(2)$^{\circ}$ }

\end{table*}

We have also performed powder X-ray diffraction on ground crystals for
$100 \le T \le 670$ and found no crystallographic changes outside of
thermal expansion and a small decrease in the angle $\beta$.  This
included temperature-dependent powder X-ray diffraction data
collected at 300, 328, 350, 380, 450, and 670 K using a Bruker D8
Advance powder X-ray diffractometer operating at 40 kV and 30 mA with
Cu K$\alpha$ radiation ($\lambda = 1.54184$ \AA ) equipped with a
temperature stage and a LYNXEYE XE detector. Data were collected in
the 2$\theta$ range of $10-80^{\circ}$ with a step size of
$0.01^{\circ}$. Rietveld refinements were performed using the TOPAS5
software package. The scale factor, background (Chebyshev function
with eight terms), and unit cell were independently refined. No
crystallographic changes were found outside of thermal expansion and a
small decrease in the $\beta$ angle.

Magnetic measurements carried out in a SQUID magnetometer revealed
behavior indistinguishable from previously grown vapor transport
crystals\cite{mendez2015}. Previous neutron scattering investigations
of the magnetic structure of Fe$_3$Ga$_4$ in polycrystalline samples
were inconclusive\cite{duijnthesis,alkanani1,alkanani2}. A single
crystal with a mass of 24 mg was chosen for magnetic structure
determination via neutron scattering using the HB-3A four-circle
diffractometer at the High Flux Isotope Reactor of Oak Ridge National
Laboratory employing a wavelength of 1.546 \AA. The sample was placed
in a closed-cycle refrigerator allowing exploration for $5 \le T \le
450$ K. We established the FM order parameter by scanning the (111)
and (201) Bragg reflections over the 5 to 450 K range. Wave-vector
scans along the (2 0 $L$) direction with $0.4 \leq L \leq 1.1$ were
collected at a large number of temperatures to characterize the
magnetic phase diagram. In addition, a large region of reciprocal
space was scanned that included all of the accessible nuclear
reflections as well as a large number of peaks associated with the AFM
ordering.  Altogether, 94 peaks were used for a full refinement of the
FM state at 5 K and 154 $q$-scans taken at 100 K were employed in
determining the AFM state. Refinement of both the nuclear and magnetic
structures was performed using the FULLPROF program
suite10\cite{fullprof}. The correlation length for the magnetic
scattering in the incommensurate spin density wave phase ($\xi$) was
determined via a fit of the data by a Lorentzian function convoluted
with the instrumental resolution.

The neutron results are presented in Tables S2 and S3 which delineate
both magnitude and orientation of the magnetic moments in the
ferromagnetic low temperature state and the intermediate temperature
spin density wave state.

\begin{table}
\begin{tabular}{|c|c|c|c|c|c|} 
\hline \multicolumn{1}{|c|}{\bf Atom} &
\multicolumn{3}{c|}{\textbf {Atomic Coordinates}} & \multicolumn{2}{c|}
{\textbf {Moment($\mu_B$)}} \\
{} &\multicolumn{1}{c}{\textbf{ \hspace{5pt} a\hspace{5pt} }}
 & \multicolumn{1}{c}{\textbf{ \hspace{5pt} b\hspace{5pt} }}
&\multicolumn{1}{c|}{\textbf{ \hspace{5pt} c\hspace{5pt} }}
& \multicolumn{1}{c}{\hspace{20pt}}{\textbf{$m_c$}}{\hspace{20pt}}
&\multicolumn{1}{c|}{\textbf{${m}^*$}}\\ 
\hline
 Fe1&0&0& 0& \hspace{0pt}1.54(28)\hspace{0pt}& 1.75\\ 
Fe2&0.4935(6) & 0 & 0.3096(7) &\hspace{0pt}1.20(22)\hspace{0pt}& 1.48\\
Fe3&0.2240(7) & 0 & 0.6376(8)&\hspace{0pt}1.24(21) \hspace{0pt}& 1.16\\ 
Fe4&\hspace{0pt}0.1352(5)\hspace{0pt} &\hspace{0pt} 0.2019(5)\hspace{0pt} 
& 0.8481(6)\hspace{0pt}& \hspace{0pt}1.49(12)\hspace{0pt}& 1.33\\
\hline\multicolumn{4}{|c|}{Average Moment}&  1.37(22)
& 1.36 \\ \hline
\end{tabular}
\caption{\label{table:FMstructure} Magnetic structure data for the low
temperature ferromagnetic state: magnetic moments determined from the
refinement of the neutron scattering data at 5 K for the four
in-equivalent crystallographic Fe sites. Moments were constrained to
lie along the c-axis. Magnetic moments from M$\ddot{o}$ssbauer
measurements taken from Ref.~\cite{Mossbauer1986} with permission of
the journal are indicated with an $*$.}

\end{table}

\begin{table*}[hbt]\centering
\begin{tabular}{|c|c|c|c|c|c|c|c|c|c|c|c|c|c|c|}
\hline \multicolumn{1}{|c|}{\bf Atom} &\multicolumn{3}{c|}{\textbf {Atomic Coord.}} &
\multicolumn{3}{c|}{\textbf {Magnetic Moments}} & \multicolumn{4}{c|}
{\textbf {$m_{mean}$($\mu_B$)}} & \multicolumn{4}{c|}
{\textbf {$m_{max}$($\mu_B$)}}  \\ %
\multicolumn{1}{|c|}{\textbf{Site}}  & \multicolumn{1}{c}{\textbf{ \centering
a}}
 & \multicolumn{1}{c}{\textbf{  \centering b }}
&\multicolumn{1}{c|}{\textbf{ \centering  c }} &\multicolumn{1}{c}{\textbf{ \centering
a}}
 & \multicolumn{1}{c}{\textbf{  \centering b }}
&\multicolumn{1}{c|}{\textbf{ \centering  c }} 
& \multicolumn{1}{c}{\centering \textbf{ a}} & \multicolumn{1}{c}{\textbf{ b}}
&\multicolumn{1}{c}{\centering \textbf{ c}}&\multicolumn{1}{c|}{ $ |m|$
}& \multicolumn{1}{c}{\centering \textbf{ a}} & \multicolumn{1}{c}{\textbf{ b}}
&\multicolumn{1}{c}{\centering \textbf{ c}}&\multicolumn{1}{c|}{$ |m|$
}\\
\hline \centering
1 &0&0& 0& -2.31(6) & 0 & 0.58(2) &1.47(4)&0&0.37(1)&1.51(4)&2.31(6)&0&0.58(1)
&2.38(6)\\
2 &0.4935(6) & 0 & 0.3096(7)& -1.55(4) & 0 & 0.38(1)&1.02(2)&0&0.25(1)&1.05(4)&1.61(4)&0&0.40(1)
&1.65(4)\\
3 &0.5065(6) & 0 & 0.6904(7)& -0.96(3) &0 & 0.24(1)&1.02(2)&0&0.25(1)&1.05(4)&1.61(4)&0&0.40(1)&1.65(4)\\
4 &0.2240(7) & 0 & 0.6376(8)& -0.95(3) & 0 & 0.24(2)&1.16(3)&0&0.29(1)&1.20(5)&1.83(4)&0&0.46(1)
&1.88(5)\\
5 &0.7760(7) & 0 & 0.3624(8)& -1.79(5) & 0 &0.45(2)&1.16(3)&0&0.29(1)&1.20(5)&1.83(4)&0&0.46(1)&1.88(5)\\
6(2)&0.1352(5)&0.2019(5)&0.8481(6)& -0.22(1) & 0.06(1) &-0.05(1)&1.32(2)&0.37(4)&0.33(1)&
1.41(8)&2.08(3)&0.58(6)&0.52(1)&2.22(8)\\
7(2)&0.8648(5)&0.2019(5)&0.1519(6)& -2.00(3) & -0.56(7) &0.50(1)&1.32(2)&0.37(4)&0.33(1)&
1.41(8)&2.08(3)&0.58(6)&0.52(1)&2.22(8)\\ \hline
\multicolumn{4}{|c|}{$m_{mean}$ } &0&0&0&1.24(3)&0.16(2)&0.31(1)
&1.30(4)&-&-&-&-\\
\multicolumn{4}{|c|}{$m_{max}$ } &-&-&-&-&-&-&-&2.31(6)&0.58(7)&0.58(2)
&2.38(8)\\\hline
\hline
\end{tabular}
\caption{\label{table:AFMstructuresup}Details of the magnetic
structure of the spin density wave state. Magnetic structural data for
the spin density wave state at 100 K from refinement of our neutron
scattering data. Data presented in this table include atomic
coordinates determined from structure refinement at $T$= 5 K, the
magnetic moment in a representative single primitive cell labeled
`magnetic moments', the mean magnetic moment, $m_{mean}$ averaged over
a large number of unit cells, and the amplitude, $m_{max}$ for each of
the different crystallographic directions. The symmetry reduction in
the AFM state results in in-equivalent magnetic moments on 7 different
Fe sites (6 and 7 have 2 fold symmetry).}

\end{table*}
\nocite{*}
\bibliography{Fe3Ga4_66}